# SEARCH FOR THE MAGNETOCALORIC EFFECT IN MULTIFERROICS OXIDES

M. BALLI[a], D. MATTE[a], S. JANDL[a], P. FOURNIER[a,b], M. M. GOSPODINOV[c]

[a] Regroupement Québécois sur les matériaux de pointe, Département de physique, Université de Sherbrooke, J1K 2R1, QC, Canada, mohamed.balli@usherbrooke.ca
[b] Canadian Institute for Advanced Research, Toronto, Ontario M5G 1Z8, Canada.
[c] Institute of Solid State Physics, Bulgarian Academy of Science, Sofia 1184, Bulgaria

ABSTRACT — In this paper, we report on the magnetocaloric properties of some selected multiferroic oxides, namely $HoMn_2O_5$ and $La_2(Ni,Co)MnO_6$ compounds which exhibit transition points from 10K up to almost room temperature. In order to avoid grain boundary effects and structural inhomogeneity observed frequently in polycrystalline samples, only single crystals were considered for this study.

## 1. INTRODUCTION

In recent years, multiferroic materials in which the magnetization/polarization can be controlled by an electric field/magnetic field have motivated a worldwide interest due to their potential use in spintronic applications. However, multiferroics could also be good candidates for magnetic refrigeration applications. The first advantage justifying a thorough study of their magnetocaloric effect (MCE) is their insulating character that prevents thermal losses caused by eddy currents. Moreover, the potential modification of their magnetic entropy by an electric field and their high resistance against oxidation and corrosion distinguish them from the usual alloys. Additionally, the combination of various physical properties in one single material offers the possibility to build efficient, miniaturized and multifunctional devices. For this purpose, we report here on the magnetocaloric potential of the multiferroics $HoMn_2O_5$ and $La_2(Ni,Co)MnO_6$.

## 2. RESULTS AND DISCUSSION

Semiconducting double perovskites $La_2(Ni, Co)MnO_6$ and their derivatives are actually the subject of several investigations due to their potential implementation in room-temperature electronic devices [1]. Their magnetic and structural characteristics are strongly dependent on synthesis conditions [2]. Both $La_2CoMnO_6$ and $La_2NiMnO_6$ are ferromagnets. $La_2(Ni, Co)MnO_6$ with high Curie points $T_C$ crystallize in the monoclinic $P2_1/n$ structure in which $Mn^{4+}$ and $Ni^{2+}$ (or $Co^{2+}$) layers alternate, while those with a low $T_C$ are Pbnm orthorhombic structures associated with disordered cations $Mn^{3+}$ and $Ni^{3+}$ (or $Co^{3+}$). Fig.1-a shows the temperature dependence of the isothermal entropy change ($\Delta S$), deduced from magnetization isotherms of a well ordered single crystal $La_2NiMnO_6$ for several magnetic fields. The $\Delta S(T)$ profiles exhibit maximum values around $T_C = 280$ K. The high value of the Curie temperature is attributed to the ferromagnetic $Ni^{2+}$ ($t^6_{2g} e^2_g$)-O (2p)-$Mn^{4+}$ ($t^3_{2g} e^0_g$) superexchange interactions. For magnetic field changes of $\Delta B = 0$-5T and 0-7T, the maximum values of $-\Delta S$ are 2J/kg K and 2.65J/kg K, respectively. For $\Delta B = 0$-5T, the obtained $-\Delta S_{max}$ is only 20 % of that of gadolinium (about 10J/kg K for 0-5T). However, the broadening of the $\Delta S(T)$ profile leads to a large refrigerant capacity ($RC = \int_{FWHM} \Delta S(T) dT$ where FWHM is the full width at half maximum). Inset of Fig.1-a shows the variation of RC of the single crystal $La_2NiMnO_6$ with magnetic field. The value of RC for a field change of 0-5T is about 206J/kg which is about 85 % of that exhibited by the giant magnetocaloric compound $Gd_5Si_2Ge_2$ [3]. Mean-field theory-based calculations (MFT) were also performed. The calculated entropy change as a function of temperature is compared with experimental results in Fig.1-a. As can be seen in Fig.1-a, $\Delta S$ (and the magnetization not shown here) is well modelled confirming the localized magnetism and semiconducting characters of the ordered ferromagnet $La_2MnNiO_6$. The observed difference between experiment and MFT data can be attributed to the fact that numerous parameters such as domain effects at 0 T are neglected in MFT. Sample imperfections could contribute also for this deviation. Besides, such calculation is of great importance in the AMR cycle modelling process. It is worth noting that under 7T, the maximum entropy change of $La_2MnNiO_6$ is only 8.3 % of the theoretical limit (31.95J/kg K) which can be deduced from MFT calculations. This means that high levels of MCE can be reached under intense magnetic fields. However, the reasons behind the low MCE in $La_2MnNiO_6$ and how to improve it must be explored in details. This will be the subject of a future investigation. Despite the observed low MCE, $La_2MnNiO_6$ presents a large refrigerant capacity, as shown above. Additionally, by changing the preparation conditions of $La_2(Ni, Co)MnO_6$ phases [2], their $T_C$ can be largely shifted to low temperatures [4, 5] and then a large working magnetocaloric temperature range can be covered as shown in Fig.1-b (inset). Raman spectra and effective magnetic moment data (note shown here) have confirmed the disordered nature of the single crystal $La_2CoMnO_6$ with $Co^{3+}/Mn^{3+}$ oxidation state, which explain its low $T_C$ (170K). So, different $La_2(Ni, Co)MnO_6$ can be combined in order to form a composite material working in the temperature range limited by their $T_C$. As an example, a composite refrigerant based on the ordered $La_2NiMnO_6$ ($T_C$ = 280K) and disordered $La_2CoMnO_6$ ($T_C$ = 170K) is simulated. The optimal mass ratio of each constituent was calculated numerically and found to vary slightly with magnetic field (about 24 % for $La_2CoMnO_6$ and 76 % for $La_2NiMnO_6$). The entropy change $\Delta S_{comp}$ of the composite is reported in Fig.1-b under different magnetic fields. $\Delta S_{comp}$ remains approximately constant over the whole considered temperature range (170-280K), which is a preferred situation for the active magnetic refrigeration (AMR) and Ericsson thermodynamic cycles.

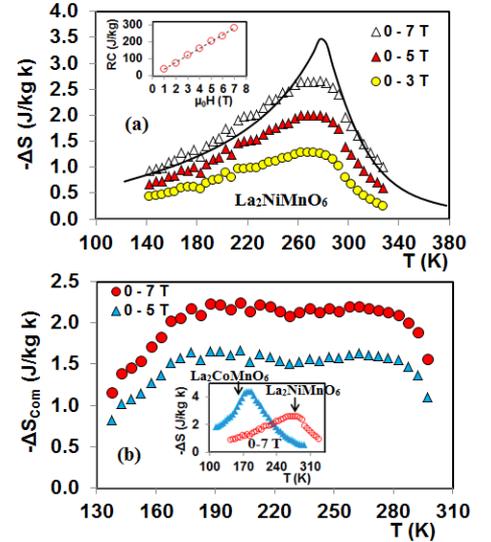

Fig. 1. (a) Isothermal entropy change of the ordered $La_2NiMnO_6$ measured in different magnetic fields. MFT data (solid line) are also given for 7 T. Inset: RC as a function of the magnetic field. (b) Isothermal entropy change for a composite material constituted of $La_2NiMnO_6$ and $La_2CoMnO_6$ compounds under 5 and 7T. Inset: individual entropy changes of both $La_2NiMnO_6$ and $La_2CoMnO_6$.

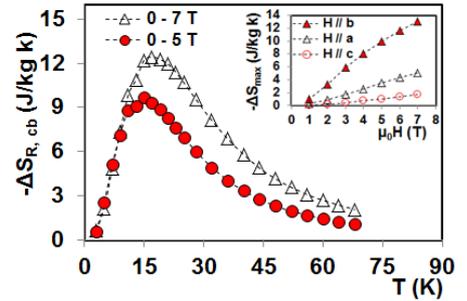

Fig. 2. Rotating MCE resulting from the rotation of $HoMn_2O_5$ within the cb plane. Inset: magnetic field dependence of the maximum isothermal entropy change for H//*b*, H//*a* and H//*c*.

For magnetic cooling application at low temperature, the multiferroics $RMnO_3$ and $RMn_2O_5$ (R = rare earth) based compounds seem to be of potential interest, since the ordering of the rare earth magnetic moments in this temperature range results in a large variation of the magnetization leading consequently to a large MCE. In this paper we focus on the orthorhombic compound $RMn_2O_5$ (R=Ho) [6], which presents fascinating magnetic and magnetocaloric behaviours. The compound $HoMn_2O_5$ was found to present dominant antiferromagnetic interactions with a gigantic magneto-crystalline anisotropy where the easy and hard directions are along the *b*, and *c* crystallographic axes respectively. A giant entropy change associated with $Ho^{3+}$ magnetic moments ordering is observed close to 10K, when the magnetic field is applied along the easy axis *b* (Fig.2 inset). Under 7T, the maximum entropy change is 13.1J/kg K for the b axis while it is only 5J/kg K for the *a* axis and negligible when the field is applied along the hard axis *c*. The refrigerant capacity reaches 334J/kg which is much higher than that of some of the best intermetallics [3]. On the other hand, since the MCE presents a giant anisotropy between the *b* and *c* axes (Fig.2 inset) in $HoMn_2O_5$, a large MCE can be also obtained by rotating $HoMn_2O_5$ within the *cb* plane (*c* is the hard direction) in a constant magnetic field instead of moving it in and out of the magnetic field region. It is worth noting that the vast majority of studies reported in the past concerns mainly the conventional MCE resulting from the change in the magnetic order. However, the anisotropy-induced rotating magnetocaloric effect can open new ways for the conception of compact, simplified and efficient rotary magnetic refrigerators [7]. By considering initially *c* direction parallel to the magnetic field, the resulting entropy change when the single crystal $HoMn_2O_5$ is rotated in the *cb* plane by 90° (H//*b*), can be expressed as: $\Delta S_{R, cb} = \Delta S (H//b) - \Delta S (H//c)$. $\Delta S_{R, cb}$ (T) is given in Fig.2 for several magnetic fields. For 5T and 7T, the maximum of $-\Delta S_{R, cb}$ is 10J/kg K and 12.43J/kg K, respectively. Based on Fig.2 data, $HoMn_2O_5$ reveals a large rotating MCE near 10K. Consequently, as we proposed for example in Fig.3, the liquefaction of hydrogen and helium can be performed by rotating continuously around 10K a regenerator constituted of $HoMn_2O_5$ single crystal blocks (or similar materials) in a constant magnetic field. Besides, such concept could be exploited in room-temperature applications [8].

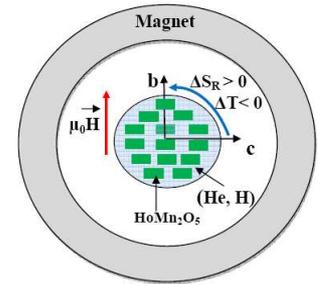

Fig. 3. Implementation of the rotating MCE for the liquifacting of helium and hydrogen (for example) using $HoMn_2O_5$ as a refrigerant.